\documentclass[]{spie}  %>>> use for US letter paper
\usepackage[]{graphicx}

\title{Laue lenses for hard X--/soft \textbf{\Huge$\gamma$}--rays: new prototype results}

\author{E.~Virgilli\supit{a}, F.~Frontera\supit{a}, V.~Valsan\supit{a,b}, V.~Liccardo\supit{a,b},
V.~Carassiti\supit{c}, F.~Evangelisti\supit{c}, S.~Squerzanti\supit{c} 
\skiplinehalf
\supit{a} \small\textit{Physics Department, University of Ferrara, Via Saragat 1, 44122 Ferrara, Italy};\\
\supit{b} \small\textit{Universit\'e de Nice Sophia-Antipolis, Parc Valrose, 06108 Nice Cedex 2, France};\\
\supit{c} \small\textit{INFN, Section of Ferrara, Via Saragat 1, 44122 Ferrara, Italy}.
}

\authorinfo{Further author information: (Send correspondence to:)\\
E-mail: virgilli@fe.infn.it, frontera@fe.infn.it, valsan@fe.infn.it, liccardo@fe.infn.it
}

\begin{document} 

\maketitle

%%%%%%%%%%%%%%%%%%%%%%%%%%%%%%%%%%%%%%%%%%%%%%%%%%%%%%%%%%%%% 
\begin{abstract}

We present the results obtained with the new Laue lens prototype built in the LARIX facility in the 
Physics Department of University of Ferrara. 
Following the results of the first prototype presented at the SPIE conference in Marseille, 
and also thanks to the methods adopted for improving the prototype (SPIE conference 
in San Diego, Ferrari et al. 2009) here we present the results of the new prototype with improved 
performances in terms of point spread function (PSF) and spectral response.

\end{abstract}


%%%%%%%%%%%%%%%%%%%%%%%%%%%%%%%%%%%%%%%%%%%%%%%%%%%%%%%%%%%%%
\section{INTRODUCTION}
\label{sec:intro}

Here we report on the status of the HAXTEL project (HArd X-ray TELescope) devoted to 
developing the technology for building broad energy passband (70/100--600 keV) Laue
lenses, highlighting test results of the new lens prototype.

We propose a Laue lens as a new focusing instrument in the soft gamma--ray band (for an 
exhaustive review of Laue lenses see Ref.~[\citenum{Frontera11}]).
Laue lenses principle exploit the interference between the periodic nature of
the electromagnetic radiation and the periodic atomic displacement as that typical of 
crystal lattice sructure.
In a Laue lens, photons pass through the full crystal, using its entire volume 
for interacting coherently. In order to be diffracted, incoming gamma--ray radiation 
must satisfy the Bragg condition:

\begin{equation}
2 d_{hkl} \sin \theta_B  = n\frac{hc}{E}
\label{e:bragg}
\end{equation}

where d$_{hkl}$ is the spacing between the lattice planes (defined by the Miller indices $(hkl)$), $n$ is 
the diffraction order, $hc = 12.4$ keV $\cdot$ $\AA$,  and $E$ is the diffracted energy at a given Bragg 
angle $\theta_B$.

A Laue lens for astrophysical applications is made of a number of crystals, in 
transmission configuration (Laue geometry), placed so as to focus the incident radiation onto a common focal spot. 
A convenient way to visualize the geometry of a crystal lens is to consider it as a part of a sphere, 
covered with crystal tiles having their diffracting planes perpendicular to the tangent plane to the sphere. 
The focal spot position lying on the lens axis defines the so called focal length of the lens $f$ which 
is at a distance $f = R/2$ from the spherical apex, $R$ being  its curvature radius. 

\section{Crystal properties and tile selection} 
\label{sec:sel}

The tiles used for building the prototype are mosaic crystals made of Copper with a cross section of 
$15 \times 15$~mm$^2$ and 3~mm thick. The lattice planes used for the diffraction are 
those related to the Miller indices (111). 

Mosaic crystals are mainly described by mosaicity $\beta$, crystallite size $t_0$ and effective 
thickness $T$. 
The mosaicity is an intrinsic property of the single tile and in principle is the same for 
each crystal, given that each of them comes from the same ingot. Imperfection during the cutting
process and at microscopic level can give rise to a different set of mosaicities.
The best cristallite size that provide the best reflectivity should be at most of the order of ten of $\mu$m.
Unfortunately also Cu (111) microcrystals beyond 200 micron have been reported [\citenum{Barriere09a}]. 

We estimated  the parameters of each crystal tile by using the measured reflectivity and comparing it
with the expected reflectivity function
(see Refs.~[\citenum{Zachariasen}] and [\citenum{Frontera11}] and references therein), after taking into account the
beam divergence.

Indeed, due to the beam divergence, the crystal is hit on its surface with different Bragg angles
and this affects the response function which is broader than that expected as a result of the crystal 
mosaicity.
If one ignores the correction for the divergence effect, mosaicity values higher than 6 arcmin 
and crystallite sizes from 200 to 400~$\mu$m are derived.
By correcting for the divergence effect, significant lower parameter values are obtained 
($\mu$ = 2.0--3.0 arcmin; t$_0$ = 30--70 micron). 

\begin{figure}[!h]
 \begin{center}
\includegraphics[scale=0.3]{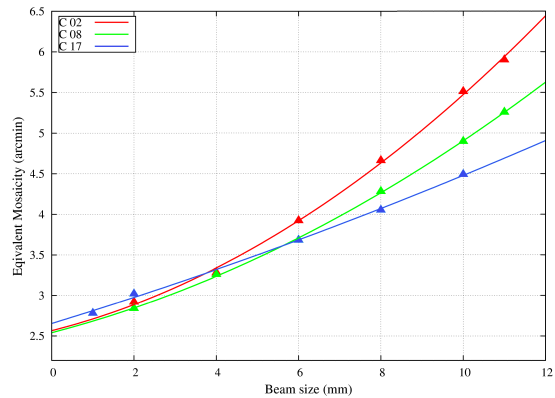}
\caption{The $equivalent~mosaicity$ as a function of the beam size for a subset of crystal. Superimposed to the 
experimental data, a polynomial function was used to fit and extrapolate the real value of the mosaicity, at 
beam size equal to zero.}
\label{fig:divergence}
\end{center}
\end{figure}

The divergence--corrected results were also tested with an experimental procedure. For a subset of
crystal tiles, lower and lower beam divergence values, obtained by decreasing the beam size 
from 10$\times$10 mm$^2$ to 2$\times$2 mm$^2$, were obtained and the corresponding
reflectivity measured. Without taking into account the correction for divergence, we derived
the gross mosaicity  (called $equivalent~mosaicity$) dependence on beam divergence shown
in Fig.~\ref{fig:divergence}.
As can be clearly seen, as the divergence converges to zero (beam size 
approaching to $\sim$ 0), the equivalent mosaicity reaches the crystal mosaicity. 
With this method, for the subset of 3 chosen crystals (No. 2, 8, 17), we estimated a crystal
mosaicity of 2.57, 2.55 and 2.68 arcmin, respectively, These values are in good agreement with the 
that estimated with the divergence corrected fitting procedure (2.72, 2.48 and 2.84 arcmin). 

\section{The Lens assembly technique and its development facility} 
\label{sec:tec}

Details of the lens assembling steps have already been reported 
[\citenum{Frontera08,Loffredo05,Ferrari09}]. In short, the adopted lens assembling 
technique is based on the use of a counter-mask provided with holes, 
two for each crystal tile. Each tile is placed on the counter-mask by means of two pins, steadily 
glued to the tile, that are inserted in the counter-mask holes. The pin axis direction and the lattice 
planes of each crystal tile have to be exactly coincident. 

The hole direction constrains the energy of the photons diffracted by the tile, while the 
relative position of two holes for each crystal in the counter-mask establishes the azimuthal orientation 
of the mean crystal lattice plane, that has to be orthogonal to the lens axis. 
Depending on the counter-mask shape, and mainly on the direction of the hole axes,
the desired geometry of lens can be obtained. In the case of a lens for space astronomy, 
the hole axes have to be all directed toward the center of curvature of the lens.
In the case of the developed prototype model, the hole axis is set parallel to that of the lens axis. 
This choice has been made to illuminate the entire lens with the available polychromatic hard X-ray source 
placed at 6 m far from the lens.

Once the average direction of the chosen lattice planes (111) of each Cu mosaic crystal tile has been 
determined, the two pins are inserted in a pin-holder which is preliminary made parallel to the hard X-ray beam.
Then the pins are glued to the crystal tile (see section \ref{sec:pinholder}). 

%A drilled counter-mask is used to accommodate the glued crystal, where each  
%couple of holes is directed towards to the center (section \ref{sec:counter-mask}). 
%The tolerance for the hole direction manufactoring with respect to the counter-mask axis is 20 arcsec. 

Once all the crystals are positioned on the counter-mask, a carbon fiber frame is placed 
above the counter-mask/crystal 
structure and glued to the entire set of crystals. The lens frame, along with the crystals, is thus
separated from the counter-mask and the glued pins are subsequently unscrewed from an aluminum glued 
head.  

The lens assembly  apparatus is installed in the LArge Italian X-ray Facility (LARIX)
located in the Physics Department of Ferrara. In addition to a set of devices employed 
for the specific project, the facility includes an X-ray generator with a fine focus of 0.4 mm 
radius with a maximum voltage of 150 kV and a maximum power of $\sim$ 200 W. The photons coming 
out from the X-ray tube are directed towards a collimator aperture which can be remotely adjusted 
in two orthogonal directions for beam size optimization. The lens building phase and performance test
are performed by means of two detectors: an X-ray imaging detector with spatial resolution of 
300 $\mu$m and a cooled HPGe spectrometer with a 200 eV spectral resolution. Both are located
on a rail and can be moved back and forth along the beam axis. They are  used to collect 
direct photons and those diffracted from the crystal tiles.

\section{Improvements with the new prototype}
\label{sec:impr}

After the development of the first prototype described in Ref.~[\citenum{Frontera08}], new  
improvements and tests have been performed in terms 
of assembling technology and of knowledge of the error budget introduced by each single step 
of the entire assembly process. A description of the improvements and changes have concerned the following topics. 
 
\subsection{Alignment of the pin-holder}
\label{sec:pinholder}
\begin{figure}[!h]
 \begin{center}
\includegraphics[scale=0.55]{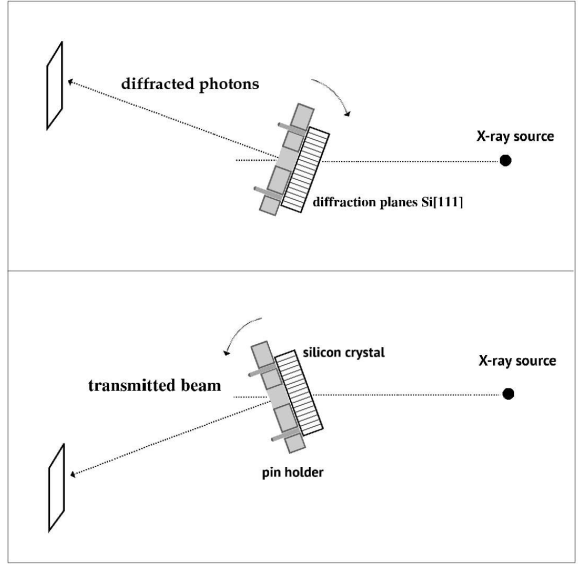}
\caption{Sketch of the pinholder alignment which is done in both horizontal and vertical direction using 
the perfect silicon crystal in Laue configuration.}
\label{fig:pinholder}
\end{center}
\end{figure}

The pinholder 
direction is preliminary made parallel to the X-ray beam axis and thus to the direction of the average 
crystalline lattice planes by means of a Silicon perfect crystal which is mechanically coupled with the pinholder.
The alignment procedure for silicon is the same employed for each copper crystal tile by means of Laue diffraction
(see Fig. \ref{fig:pinholder} for a schematic view of the alignment) and it is performed using the lattice 
planes (220), which are orthogonal to the crystal external surface.

\begin{figure}[!h]
 \begin{center}
\includegraphics[scale=0.35]{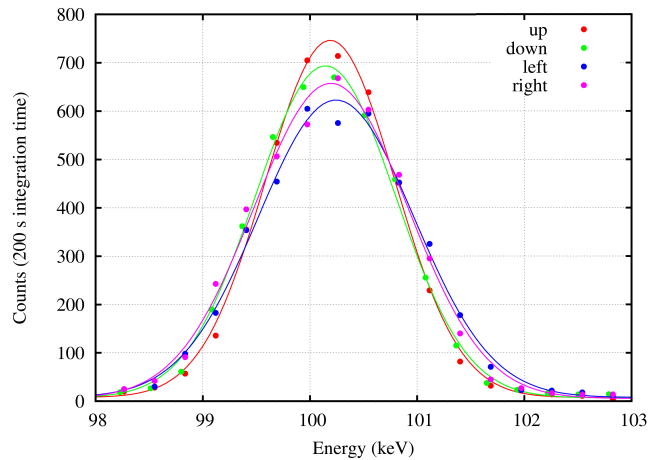}
\caption{Diffraction spectra of silicon crystal showing up($red$), down($green$), left($blue$), right($purple$) diffraction 
with respect to the lattice planes (220). A gaussian function for best fit is superposed to each 
spectra, to determine the peak energy.}
\label{fig:silicon_diff}
\end{center}
\end{figure}

The chosen lattice plane direction of the Silicon is obtained when exactly the same energy 
(typically 100 keV) is diffracted with the crystal rotated at left and right around the crystal vertical axis. 
The verticality of the lattice planes is established from the image position of the crystal obtained 
when it is rotated at left and right direction. The diffracted spectrum when the crystal is rotated upward
and downward around its horizontal axis confirms the chosen lattice plane verticality.  
%
%Similarly, the diffraction was performed rotating upward and 
%downward the silicon crystal around its orizontal axis. 
%
Figure \ref{fig:silicon_diff} shows the results 
of the achieved alignment: all the diffracted spectra of the polychromatic beam have a Gaussian profile
with a full width at half maximum (fwhm) consistent with the spectral resolution of the detector and a 
centroid energy consistent with each other (E$_{up}$ = 100.19 keV, E$_{down}$ = 100.14 keV, 
E$_{left}$ = 100.24 keV, E$_{right}$ = 100.20 keV).

Being the silicon crystal mechanically coupled with the pinholder, and being the pinholder alignment 
made within a tolerance better than 30 arcsec, which is related to the working tolerance of the drilled holes, 
the uncertainty in the lattice plane determination is mainly due to the mechanical coupling 
between silicon crystal and pinholder.

\subsection{Pin shape and manufacturing}

As a reference for the beam axis and thus of the chosen lattice plane direction, 2 pins are glued on each
crystal. The gluing phase is one of the most 
delicate part of the process because of the subsequent separation of the glued pins from the pinholder 
and of the final step of removing crystals from the counter-mask. 

As already reported [\citenum{Ferrari09}], different pin shapes have been tested. For building the 
first prototype, we used the cylindrical pin geometry.
Later, to decrease the friction between pins and pinholder during the separation phase, instead of using
two cylindrical pins, we tested a single pin with conical shape and elliptical cross-section for each crystal tile . 
This geometry did not give positive results. The post-gluing tests showed a systematic 
mismatch between left and right diffracted energies, even greater than 2 keV with respect 
to the pre-gluing energies. For comparison, the same pre/post gluing energies obtained with the 
cylindrical pins give a $\Delta$E $\leq$ 0.3 keV. We concluded that
the lower strength of the crystal--pin coupling, due to the smaller gluing surface is not 
compensated by the lower friction between pin and pinholder.

\subsection{New counter-mask and carbon fiber support}
\label{sec:counter-mask}

\begin{figure}[!h]
\begin{center}
\includegraphics[scale=0.18]{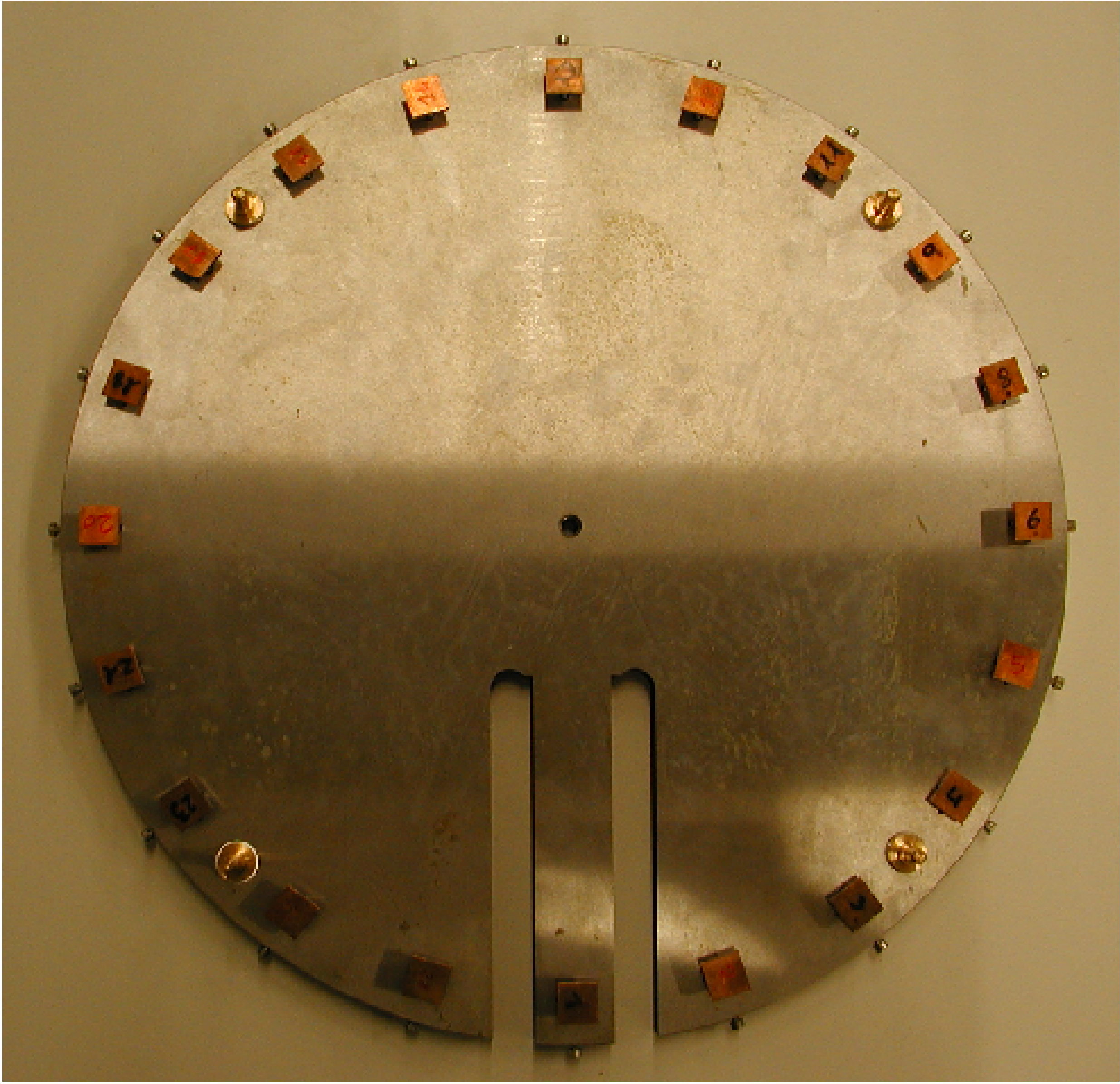}
\includegraphics[scale=0.18]{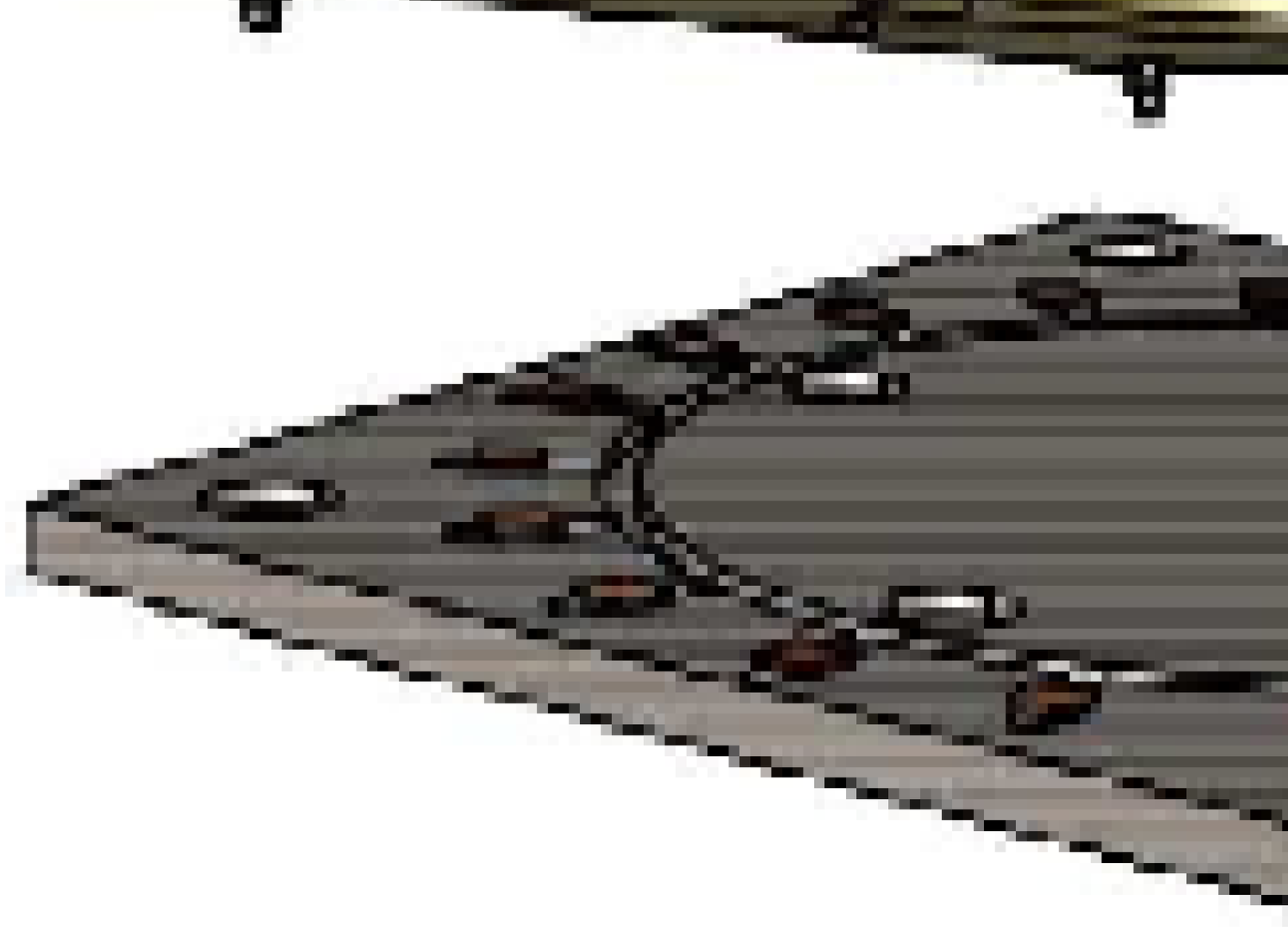}
\caption{The counter-mask with the 20 crystals placed, and the carbon fiber support.}
\label{fig:counter-mask}
\end{center}
\end{figure}

The new lens prototype consists of 20 mosaic crystals of Copper (111), with a mosaic spread 
of $\sim$2 to $\sim$3 arcmin. The crystal tiles are arranged in a ring configuration with a 
diameter of 36 cm, like the first one  [\citenum{Frontera08}]. The lens is assembled by means of a 
counter-mask (Fig. \ref{fig:counter-mask}) in which 
20 pairs of holes are drilled with a radial orientation. Each crystal tile is positioned
on the counter-mask by inserting its two cylindrical pins in a pair of holes.
To minimize the sources of errors, once all the
crystals are positioned on the counter-mask, the two pins are tightened in the holes by means of 
lateral screws, thus 
reducing their movement and then the misalignment of the tiles, in particular when
the lens frame is being glued. 

A thin plate (see Fig. \ref{fig:counter-mask}) 1 mm thick is used as a lens frame. It is 
made of 8 layers of carbon fiber properly oriented in order to avoid any stress or warp due to 
temperature changes.
The frame is positioned and stuck on the entire set of crystals. Then the so assembled lens
is separated from the counter-mask. The carbon structure combines an excellent efficiency of transmission 
($\sim$ 95\% at 100 keV)  with a toughness and thermal stability.

\subsection{Alignment capability}
\label{sec:counter-mask}

Given the left/right technique for the determination of the mean lattice planes direction, its accuracy 
is given by the spectral resolution of the HPGe detector (0.2--0.3 keV). It follows that, using
the relation $\Delta$$E$/$E$ = $\Delta$$\theta$/$\theta$, at 100 keV the angular accuracy in the
lattice plane direction is determined with a precision of $\Delta$$\theta$ $\simeq$ 15--20 arcsec.

On the other hand, measuring  the image misalignment as detected with the X-ray imager 
(spatial resolution of 300 $\mu$m) and taking into account the distance between the crystal and 
the detector, the minimum angular resolution is only 2 arcmin. Thus the errors introduced by 
the adopted alignment procedure are not detectable by 
the imager detector.  

We conclude that, in the case we observe a PSF (Point Spread 
Function) of the focused photons with a spread higher than 20 arcsec, this is certainly
due to the errors introduced by the mechanical separation of the lens frame 
from the counter-mask.

\section{Lens testing  and results}

Once the assembled lens is separated from the counter-mask, for its testing the lens is positioned 
on a support of the LARIX facility located at half way (about 6 m) between gamma-ray source and
focal plane detectors. These can be remotely translated back and forth along the beam axis for
finding the best  focusing position and for getting, out of focus, the image position of each crystal.
 
\begin{figure}[!h]
\begin{center}
\includegraphics[scale=0.85]{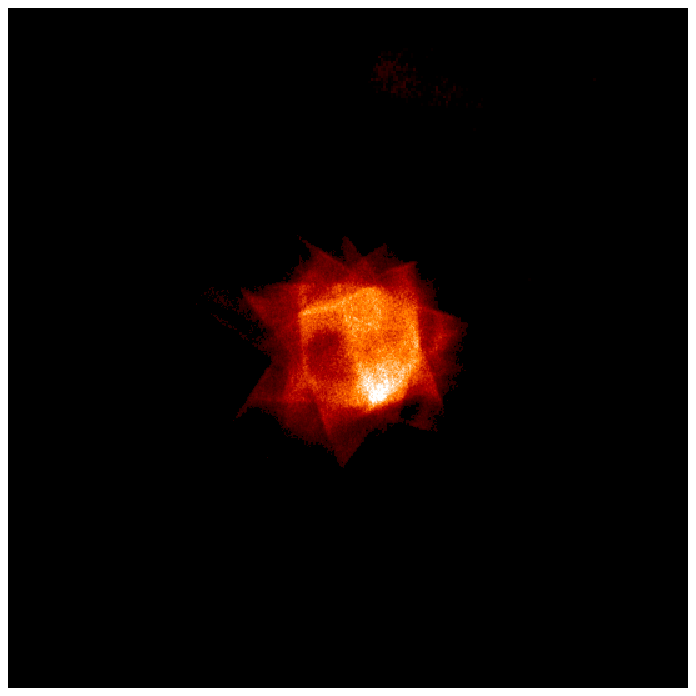}
\includegraphics[scale=0.85]{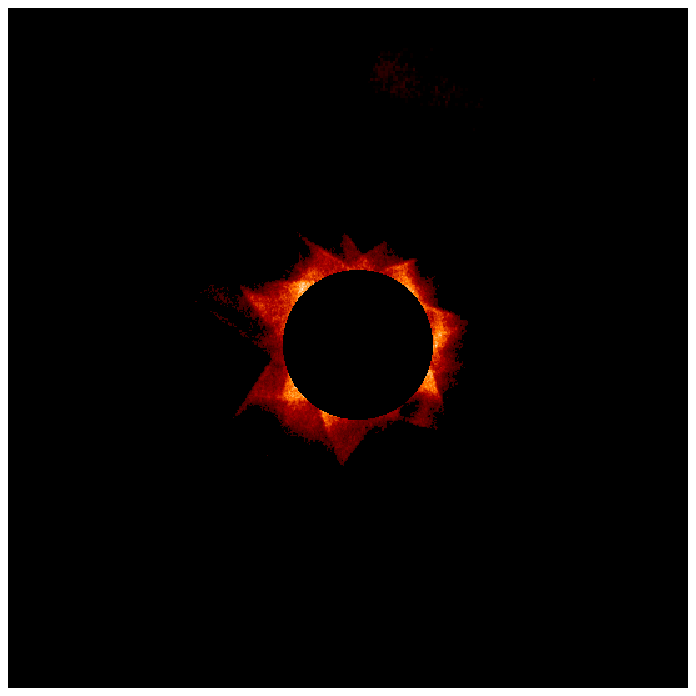}
\caption{$Left~panel$: Point spread function obtained with the developed prototype. $Right~panel$: Difference 
between the prototype PSF and the spot expected from Monte Carlo simulations assuming 
diffraction images from perfectly aligned crystals.}
\label{fig:light}
\end{center}
\end{figure}

To avoid  direct radiation to arrive on the focal plane detector, the entire inner region of the 
lens frame is covered by a lead layer 3 mm thick.  The left panel of 
Fig. \ref{fig:light} shows the first light of the developed prototype when the polychromatic 
beam irradiates the entire lens.

With Monte Carlo techniques, and accounting for the divergence of the beam as well as
the mosaicity of the crystals, the PSF, for an ideal lens with diffraction from  
perfectly aligned crystals, was derived. The comparison between the expected PSF and that
obtained with the lens prototype is shown in Fig.~\ref{fig:light} (right panel). 
The dark circular region shows the expected PSF.

\begin{figure}[!h]
\begin{center}
\includegraphics[width=11cm, height=7cm]{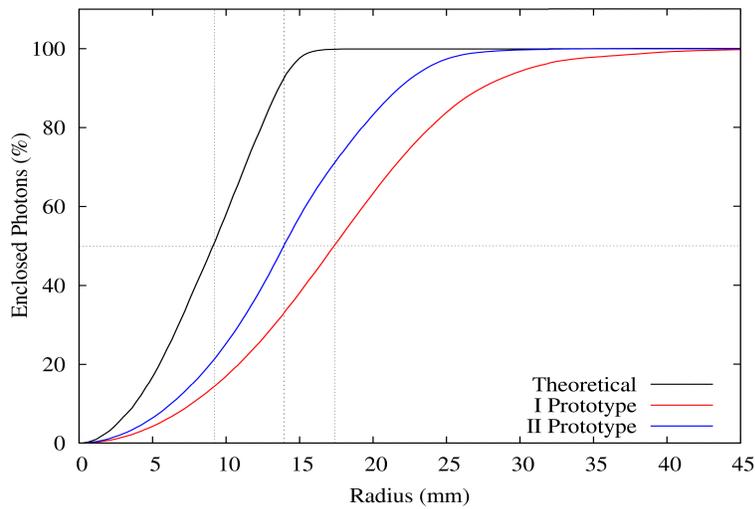}
\caption{Cumulative distribution of the focused photons along the radial distance from the 
focal point. $Black~line$ corresponds to the expected distribution in the case of a perfect alignment 
of the crystals. The $red~line$ shows the photon distribution obtained in the first prototype 
([\citenum{Frontera08}]) while the $blue~line$ shows photon distribution for the second prototype.}
\label{fig:collected}
\end{center}
\end{figure} 

Instead Fig.~\ref{fig:collected} shows 
the cumulative distribution of the number of photons collected along the radial distance from the focal point.
As can be seen, while the expected half power radius (HPR), i.e. the radius within which 50\% photons are collected,
is 9~mm, the measured HPR was 17.4 mm for the first prototype [\citenum{Frontera08}], and 13.9 mm for 
the second prototype. There is an improvement of 41.66 \% with respect to the first prototype, 
considering theoretical value of the radius(9 mm) as the target.
Thus this  represents a significant improvement with respect to the first one.
It can be also seen that, at the radius (16.00 mm) at which the saturation occurs for the expected 
cumulative distribution, the fraction of collected photons is around 0.7, with respect to the first prototype in
which the fraction was less then $60\%$.

\begin{figure}[!h]
\begin{center}
\includegraphics[scale=0.33]{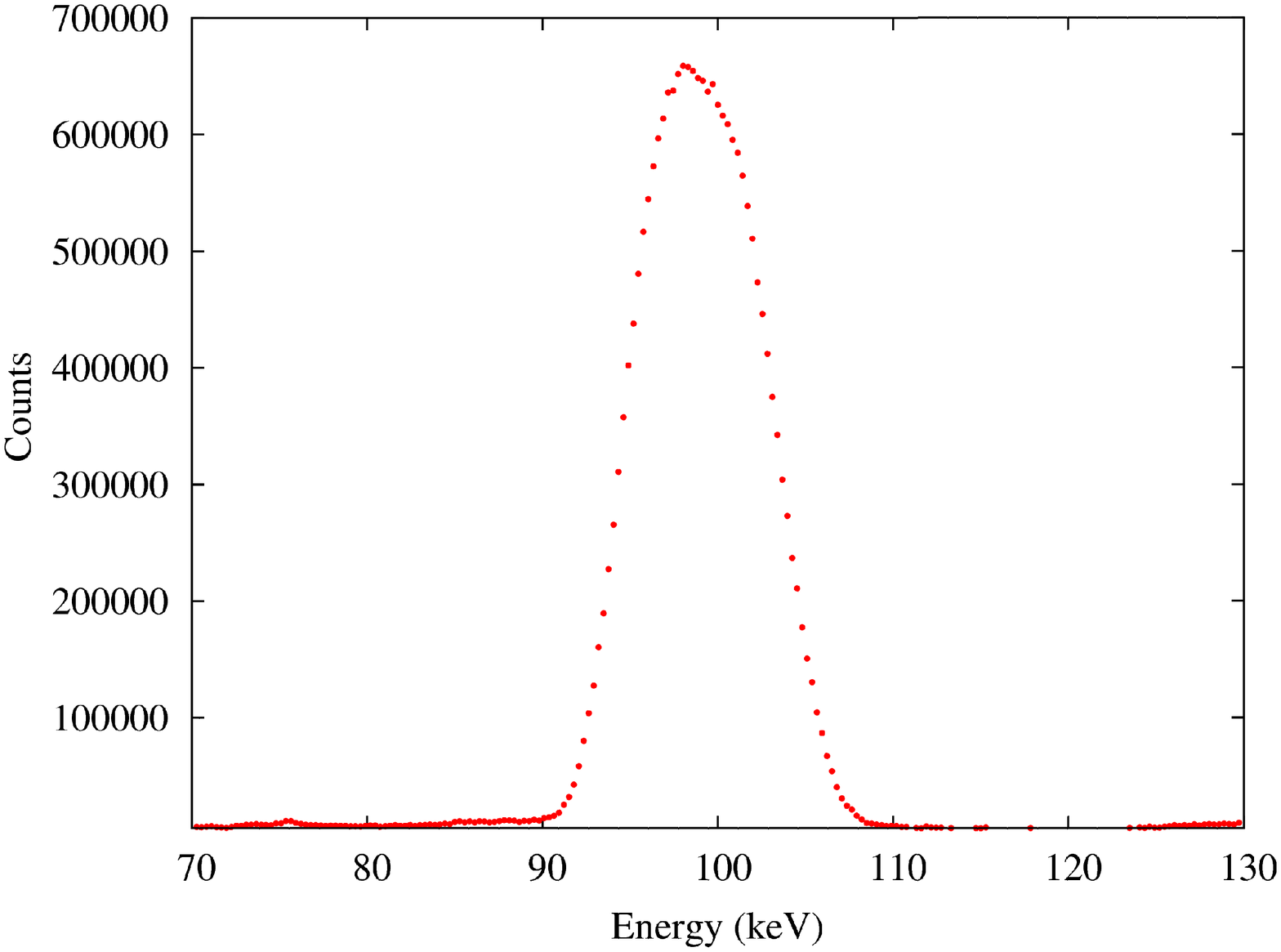}
\includegraphics[scale=0.33]{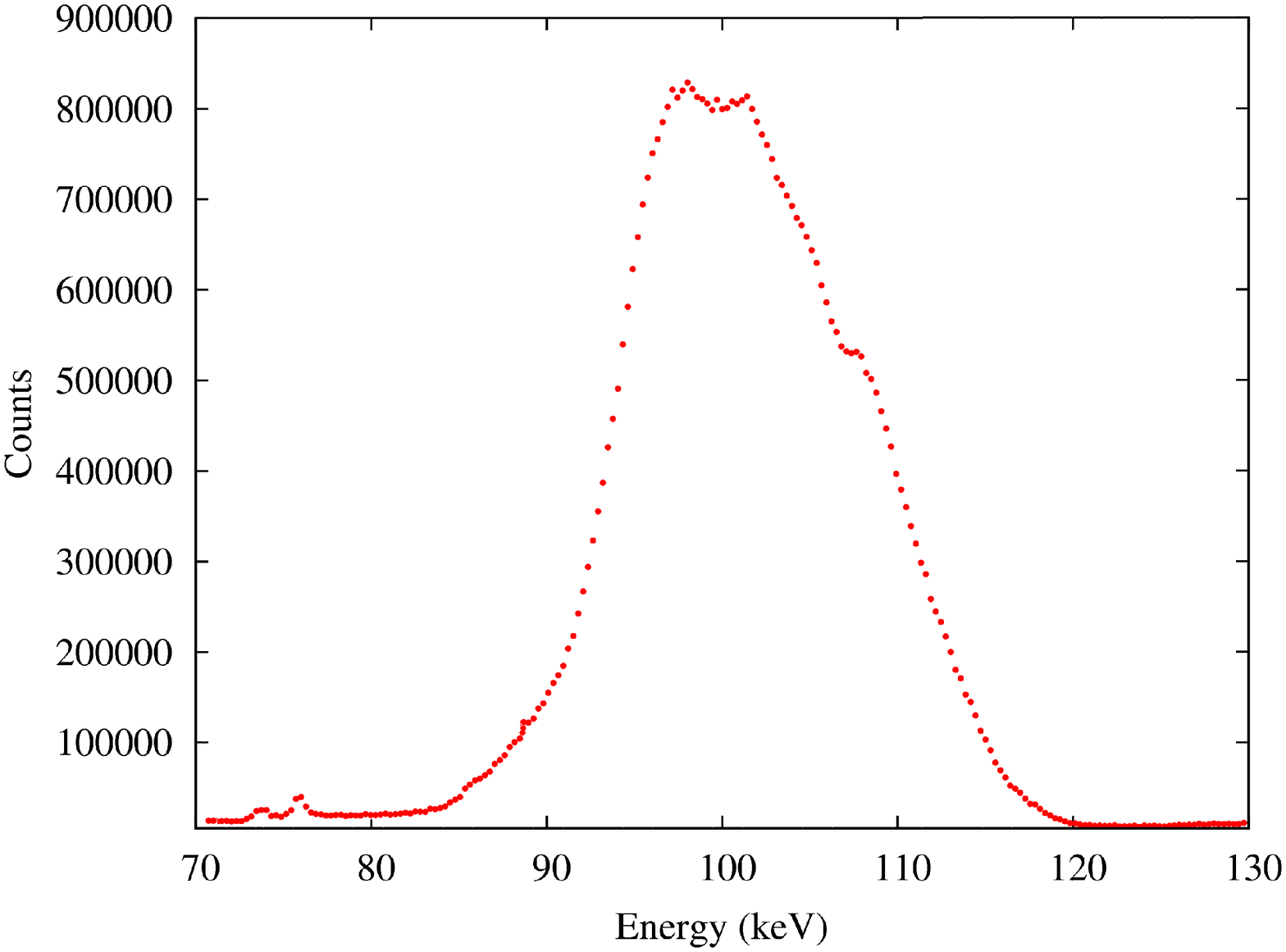}
\caption{ $Left~panel$: Spectrum of the total photons reflected on the focus.
$Right~panel$: Spectrum of the total photons reflected by the lens in the region 
in and around the focus, in the focal plane.}
\label{fig:spectrum}
\end{center}
\end{figure}

The spectral analysis of the focused beam was also performed and the results are presented 
in Fig. \ref{fig:spectrum}. The lens prototype was designed for giving a spectral 
peak at 100 keV at the focal point. In the left panel it is shown the spectrum obtained collecting 
only the photons coming in the central region of the focus, while in the right panel, it is shown
the spectrum of all the photons reflected by the lens and collected on the focal plane. 
In both cases the exposure time was 1000 seconds. As can be seen, in the central 
region the peak is at 98.89 keV. The centroid of the spectrum of the central region achieves an 
intensity level 0.8 times that of the peak spectrum of all reflected photons.

Also a thorough spectral analysis was performed for each single crystal, in order to 
estimate the misalignment distribution in terms of centroid energy diffracted from each 
crystal tile. The results are reported in Table \ref{tab:angular} in which, along with the peak  
energy and the FWHM of the spectrum diffracted by each crystal, the angular deviation of each
crystal from the theoretical one is also evaluated.

\begin{table}[!h]
\label{tab:angular}
\begin{center}
\footnotesize
\begin{tabular}{c|c|c|c||c|c|c|c}
%\multicolumn{2}{c|}{$\mu$} & \multicolumn{2}{|c}{t$_0$}\\
Crystal & Peak Energy &  FWHM & $\Delta \theta$ & Crystal & Peak Energy & FWHM & $\Delta \theta$\\
 Number & (keV) &  (keV) & (arcmin) &  Number &  (keV) & (keV) & (arcmin)\\
\hline
   1 &   102.31 & 6.53 &  -3.54 &   13 &    96.88 & 5.39 &    2.08  \\
   2 &    98.37 & 5.01 &   0.53 &   14 &   103.70 & 6.83 &   -4.99   \\
   3 &   101.47 & 5.65 &  -2.67 &   16 &   104.66 & 6.15 &   -5.99   \\
   4 &   104.45 & 7.13 &  -5.76 &   17 &   103.92 & 5.94 &   -5.22   \\
   5 &    97.09 & 4.50 &   1.87 &   18 &   105.27 & 6.15 &   -6.62    \\
   6 &    99.29 & 5.31 &  -0.41 &   20 &    99.79 & 6.67 &   -0.93   \\
   8 &   103.10 & 6.33 &  -4.36 &   21 &   103.81 & 5.17 &   -5.10   \\
   9 &   104.49 & 6.31 &  -5.81 &   23 &   104.52 & 7.50 &   -5.84   \\
  11 &    96.42 & 5.12 &   2.53 &   27 &    94.12 & 6.78 &    4.95     \\
  12 &    98.60 & 5.33 &   0.29 &   28 &    95.27 & 7.08 &    3.76    \\
\end{tabular}
\newline
\newline
\caption{\footnotesize Measured peak energy, FWHM and angular deviation for each of the crystals with 
respect to the diffracted energy of the lens.}
\end{center}
\end{table}
\normalsize

\begin{figure}[!h]
\begin{center}
\includegraphics[scale=0.4]{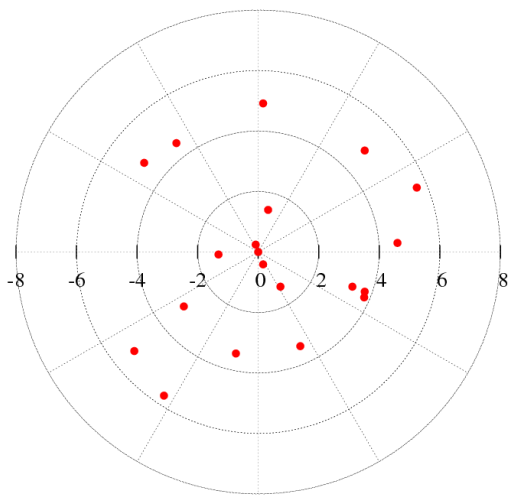}
\caption{Angular deviation of each crystal on the lens with respect to the perfect alignment.}
\label{fig:angular}
\end{center}
\end{figure}

By shielding all the lens crystals but one, and measuring the barycenter coordinate of each diffracted 
spot, also the angular misalignment was determined for each 
lens crystal. Figure \ref{fig:angular} shows the deviation occured for each crystal, from their expected perfect position.
For the total set of tiles, this deviation is within 6 arcmin (for the first prototype it was 15 arcmin). 

\section{Discussion}
We have presented the performance of a focusing lens made of 20 crystals, assembled
by in the LARIX facility of the University of Ferrara.

Compared to the first prototype, we have increased the performance in terms of imaging 
capability. The maximum angular deviation of the crystal tiles from their nominal ones
has been decreased from 15 arcmin to 6 arcmin.

The contribution of each step to the overall error budget has been evaluated. 
The alignment of the lattice planes with the gamma-ray beam axis  has been performed with 
a precision better than 30 arcsec. The pin-holder alignment with the beam axis
has been estimated to show a similar precision. 
The holes of the pin-holder are drilled parallel to the beam axis with an uncertainty of less 
than 20 arcsec. The same tolerance is in the direction of the counter-mask holes.
Considering the 6 arcmin of mismatch between real and expected positioning of the crystals, we attribute
this mismatch mainly to the mechanical insertion of the pins in the counter-mask and to the successive
separation process of the entire lens from the counter-mask.

In order to improve further the lens PSF, the only way we see is that of adopting a different
assembling technique, in which each crystal tile is directly glued on the lens frame under a gamma-ray
beam control. This is the goal of the LAUE project supported by the Italian Space Agency ASI, which is
on the way for the production of a lens petal prototype with 20 m focal length with an accuracy in the
crystal orientation better than 10 arcsec~[\citenum{Frontera11}].

\acknowledgments     
 
The authors wish to thank  C.~Guidorzi for his useful comments and discussions. We acknowledge the financial 
support by the Italian Space Agency (ASI) and also the contribution by the Italian Institute of 
Astrophysics (INAF). The design study was possible thanks to the 2002 Descartes 
Prize of the European Committee awarded to one of us (F.~Frontera).

\bibliography{prot}
\bibliographystyle{spiebib}

\end{document}